# Emission Characteristics of the Projectile Fragments at Relativistic Energy

M. K. Singh*[1], A. K. Soma[2], Ramji Pathak[1] and V. Singh[2]

1. Department of physics, Tilakdhari Postgraduate College, Jaunpur - 222002
2. Nuclear and Astroparticle Physics Lab, Department of Physics, Banaras Hindu University, Varanasi - 221005
*singhmanoj59@gmail.com

## Abstract

A projectile ($^{84}Kr_{36}$) having kinetic energy around 1 A GeV was used to expose NIKFI BR-2 emulsion target. A total of 700 inelastic events are used in the present studies on projectile fragments. The emission angle of the projectile fragments are strongly affected by charge of the other projectile fragments emitted at same time with different emission angle is observed. The angular distribution studies show symmetrical nature for lighter charge projectile fragments. The symmetrical nature decreased with the charge of projectile fragments. At ~4° of emission angle for double charge projectile fragments, the momentum transfer during interaction is similar for various target species of emulsion were observed. We also observed a small but significant amplitude peaks on both side of the big peak for almost all light charge projectile fragments having different delta angle values. It reflects that there are few percent of projectile fragments that are coming from the decay of heavy projectile fragments or any other process.

## I. Introduction

Nuclear emulsion detector is one of the oldest detector technologies and has been in use from the birth of the experimental nuclear and astroparticle physics. Fortunately, it is a unique and simple detector till today, due to very high position resolution (~1 μm) along with several unique features. Nuclear emulsion detector has 4π detection capability with hit density of 300 – 500 grains per mm, compactness of the size and large range of ionization sensitivity depends upon the nature and need of the experiment. The high resolution allows easy detection of short-lived particles like the τ lepton or charmed mesons. 4π visualization of tracks formed or produced during interaction impelled us to pursue studies on physics beyond the Standard Model. The detector technology coupled with modern scanning technologies enhances our understandings of nuclear reactions mechanism and formation of new state of matter.

## II. Experimental configuration

Nuclear emulsion is a detector composed of silver halide crystals immersed in a gelatin matrix [1,2] consisting mostly of hydrogen, carbon, nitrogen, oxygen, silver and bromine while a small percentage of sulfur and iodine are also present. In the present experiment, we have employed a stack of high sensitive NIKFI BR-2 nuclear emulsion pellicles of dimensions 9.8×9.8×0.06 cm$^3$, exposed horizontally to $^{84}Kr_{36}$ ion at a kinetic energy of around 1 GeV per nucleon. The exposure has been performed at Gesellschaft fur Schwerionenforschung (GSI) Darmstadt, Germany.

Interactions were found by along-the-track scanning technique using an oil immersion objective of 100×magnification. The beam tracks were picked up at a distance of 5mm from the edge of the plate and carefully followed until they either interacted with emulsion nuclei or escaped from any surface of the emulsion. These events have been examined and analyzed with the help of an OLUMPUS, binocular optical microscope, having total magnification of 2250X and measuring accuracy of >1μm.

In order to obtain an unbiased sample of events along-the-track scanning technique has been employed [3]. A total of 700 inelastic events produced in $^{84}$Kr- emulsion interactions were located. During event scanning, we picked up all genuine events in accordance with the event selection criteria mentioned in [4]. The interaction mean free path (λ) of $^{84}$Kr in nuclear emulsion has been determined and found to be 7.50±0.28 cm, consistent measurements were reported by DGKLMTV Collaboration (λ) 7.10±0.14 cm. In the present analysis, out of 700 there are 570 events having full fill the required criteria of further investigated.

The mean number of fully developed and well separated grains per unit length is called the grain density g. It is a measure of the rate of ionization loss. The grain density of a singly charged particle passing in the same emulsion at extreme relativistic velocity is called the minimum grain density ($g_{min}$). In this experiment, its measured value is 28±1 grains per 100 μm. The grain density of a track corresponds to a particular specific ionization but its actual value depends on the degree of development of the emulsion and the type of the emulsion used. It is therefore, necessary to introduce another quantity called normalized grain density which is defined as g* = g/$g_{min}$. Here g is the observed grain density. All charged secondaries emitted or produced in an interaction are classified in accordance with their ionization, range and velocity into the following categories:

(a) Shower tracks ($N_s$): These are freshly created newly produced charged particles with $g^* < 1.4$. These particles have relative velocity ($\beta$) > 0.7. For the case of a proton it means energy of $E_p$ > 400 MeV. They are mostly fast pions with a small admixture of Kaons and of released protons from the projectile which have undergone an interaction. These conditions ensure that showers are filtered from the fragments and knockout protons of the target.

(b) Grey tracks ($N_g$): Particles having ionization in the interval $1.4 < g^* < 6.0$ and range L > 3 mm are defined as greys. This particle has relative velocity ($\beta$) in between 0.3 - 0.7. They are generally knocked out protons of targets having energy $30 < E_p < 400$ MeV but also admixture of deuterons, tritons and some slow mesons.

(c) Black tracks ($N_b$): Particles having range L < 3 mm from interaction vertex from which they originated and $g^* > 6.0$. This corresponds to a relative velocity $\beta < 0.3$ and a proton with energy $E_p < 30$ MeV. Most of these are produced owing to evaporation of residual target nucleus. The heavily ionizing charged particles ($N_h = N_g + N_b$) are parts of the target nucleus and are also called target fragments.

(d) Projectile Fragments ($N_f$): These are the spectator parts of the projectile nucleus with charge Z ≥ 1 having velocity close to the projectile velocity. The ionization of projectile fragments (PFs) is nearly constant over a few mm of range and emitted within a highly collimated forward narrow cone of ±10° whose size depends upon the available beam energy. The forward angle is the angle whose tangent is the ratio between the average transverse momentums ($p_T$) of the projectile fragments to the longitudinal momentum ($p_L$) of the beam. Taking $p_L$ as the beam momentum itself, i.e., Angle (F) = $\tan^{-1}(p_T/p_L)$ = ~9° in this experiment. The PFs are further classified into three categories as follows:

(i) Heavy Projectile Fragments ($N_f$): PF's with charge Z ≥ 3.
(ii) Alpha Projectile Fragments ($N_\alpha$): PF's having charge Z=2.
(iii) Singly charged relativistic Projectile Fragments ($N_{z=1}$): PF's having charge Z=1.

Since, these PF's have velocities nearly equal to the initial beam velocity; their specific ionization may be used directly to estimate their charge. The experimental detail has been discussed in Ref. [5].

## II (a) Charge estimation of Projectile Fragments

When any charged particle passes through the medium, it transfers partial or total energy to the atoms or molecules of the surrounding medium due to interactions or scattering. If the transferred energy is large enough to make the outer most orbital electron free hence ionizes the medium and forms charge particle tracks. The rate of ionization depends on the square of the charge and inverse of the square of the velocity of the ionizing particle according to the Bethe-Block's formula [6]. Ionization measurements are a great help in estimating the charge of the projectile fragments. When the ionization is low, the certainty in such estimation may be large but as the grain density (i.e. number of fully developed and well separated grains per 100 μm) increases, the adjacent grain becomes unresolved even under a high magnification microscope that increases the uncertainty in the charge's estimation. In case of higher charge tracks the grains get clogged to each other to form blobs and it is not possible to count the individual grains. Therefore, different methods such as grain density, blob and hole density or gap length coefficient, mean gap length, delta rays counting, relative track width measurement for higher charge fragments, and residual range method is applicable for the projectile fragment charge close to the beam charge (36 charge unit in this case), based on the different ionization related parameters for a quantitative measurement of the rate of energy loss have been devised.

In our experiment, we used $^{84}Kr_{36}$ nuclei as a projectile having energy 1 A GeV. Projectile kinetic energy (84 GeV) is above the relativistic energy criteria. The charge of the projectile is 36 units and in case of peripheral and quasi central collisions there are chance to estimate the total charge (Q, sum of all projectile fragments charges) of the interaction/event more than 36 units due to neutron conversion into proton and minimum up to 1 unit of charge.

A single method can't be applied to estimate the charges over the entire range as every method has its own limitations [7]. We have adopted the grain density method for estimation of charge of projectile fragments having charge $Z \leq 4$. The gap length coefficient method is among the most accurate methods for determining the charge from 5 to 9 and from 10 to 19 is estimated by the delta (δ) rays (recoil electrons having kinetic energy more than 5 keV that acquire delta shape with inclination opposite to the beam direction) density measurement. The fragments having charge in between 19 to 30 have been estimated by the relative track width measurements [1] and residual range method is applicable for the fragments having charge above 30. The projectile fragments charge spectrum presented in this analysis are up to 10 unit charge and applied methods described below:

## II (a-1) Blob Density

A blob is defined as single structure or set of grains clogging to each other. The gap between two adjacent blobs is called hole. The number of blobs (holes) per 100 µm is called blob (hole) density and represented by B (H). If the ionization is small, the blob density is a good parameter [5,7,8] to resolve the charges but in case of particles with heavy charge, the blob density increases up to Z = 2 and then drops as blobs continue to coalesce into larger blobs as shown in figure 1. Thus for very large ionization, blob density method is not sensitive. Hence it can be applied over only a limited range of ionization. We have first measured the B and H for the projectile fragments which are clearly having charge Z = 1, 2, 3 and 4. The measurements, plotted as functions of $Z^2$ are shown in figure 1 (a) and (b) respectively. Therefore, B and H measurements alone can not determine the charge over the entire range of ionization. The nature of 1 (a) is well described by Landau distribution with peak at 5.71±0.21 and sigma of 3.30±0.18 where as 1 (b) is described by exponential function with slope –6.32 ± 0.28 and constant 3.27±0.03.

## II (a-2) Gap-length Coefficient Method

The distance between the two successive blobs is defined as gap length. This length is related to the ionization caused by the charged particle [8, 7]. The ratio of the total number of observed gaps to the number of gaps greater than a certain optimum value or the negative slope (*G*) of the log of frequency distribution of gap length is a measure of the grain density and is called Gap Length Coefficient (*G*). The gap-length coefficient (*G*) of the exponential is nearly proportional to the rate of the energy lose of the ionizing particle and is obtained by using the Fowler-Perkins [2] relation:

$$G = - (1/L) \ln (B/H).$$

Where, B is defined in section II (a), L is the suitable minimum chosen distance between the inside edges of developed grains bordering the gap and H is the number of gaps greater than a certain optimum value "L" normalized to the unity.

For considerably low ionization, one may also determine gap-length coefficient from blob density alone from the following relation:

$$B = G \exp(-\alpha G).$$

Here α is the mean diameter of a developed grain [3]. For projectile fragments whose charge could be estimated with this method with ±1 charge unit certainty to be up to Z = 8. We have computed G and plotted it as a function of $Z^2$ in figure 2. The shape of the curve is similar to the earlier reported curves [9].

According to Fowler [10], the optimum value of L occurs when GL ~ 2.0 for all values of G. The accuracy does not vary appreciably in the interval 1.5 < GL < 2.5. The statistical error in G can be expressed as

$$dG / G = 1 / [(N_H) \ln (B/H)]^{1/2}$$

Where, $N_H$ is the number of gaps greater than the length L and H is number of gaps greater than a certain optimum value L, normalized to unit length. We obtained the optimal value of L as 0.98 μm which is same as mentioned in Ref. [7]. Minimizing in error is done by setting

$$d/dH \, [H \log (B/H)]^{1/2} = 0$$

Solving the differential equation one find that

$$\ln(B/H) = 2.0 \text{ or } B/H = 7.4$$

The estimation of error is quite reliable as long as $(B/H) > 4$ and $N_B > 4N_H$.

The measurement of gap length coefficient was done for around 600 fragments tracks. We have taken into account all the gaps greater than one division of the microscope scale fitted in the eyepiece. The calibrated value of 1 division is equal to 0.98 μm. However, we do not find any significant change in our final results by varying the value of L, since the measurement for each track is based on counting large number of blobs and gaps and selection criteria for the charge measurement is responsible for different values of B and H.

## II (a-3) Delta Ray-Density Method

In a sensitive emulsion, a particle moving at relativistic velocity shows narrow, dense central core around the trajectory of the primary particle and number of delta rays which becomes more and numerous with the charge of primary particle. This method is suitable for particles (fragments) with $Z \geq 10$, the tracks of which virtually have no gaps, comprises in measuring the number and / or track lengths of δ shaped electrons produced by charged particle as it ionizes the substance along its track. This method is based on the fact that the energy and range distributions of delta electrons are dependent on charge (Z) of the ionizing particle [11]:

$$(d^2N/dTdx) = [(\tfrac{1}{2}) (4\pi N_A r_e^2 m_e c^2) \cdot Z^2 \, (z/A) \, (1/\beta^2) \, (F/T^2)],$$

where T is the kinetic energy of delta electrons, x is the thickness of the substance passed by the ionizing particle, A is the atomic weight of atoms in the substance, z is the charge of atoms in the substances, β = (v/c), v is the velocity of the ionizing particle, c is the velocity of light, $N_A$ is the Avogadro's number, F is the parameter dependent on the spin of the ionizing particle at relativistic velocities, it is considered to be constant, $m_e$ is the mass of electron and $r_e$ is the classical radius of electron. The second term of right hand side is equal to 0.3071 MeV.cm$^2$g$^{-1}$[11].

The number of the delta rays having kinetic energy more than 5 keV, which escaped from the parent particle, may produce recognizable delta shape (δ) tracks with three or more grains inclined against the direction of the parent particle, in this measurement. The delta rays having above criteria will contribute to the value of the delta ray density. According to the Demers and Rossi [12], at relativistic velocity, the maximum value of energy of knock-off electrons becomes large compared to any measured minimum delta ray's energy. So the number of delta rays exceeding a particular minimum energy ($W_{min}$) will becomes $N_\delta$, which is

$$N_\delta \sim (2\pi r_0^2)\,[(m_e c^2)/W_{min}]\,Z^2.$$

Here $m_e$ and c is the mass of electron and velocity of light in vacuum, respectively. Generally, we choose a fixed value of $W_{min}$ for an experiment. In the above equation right hand side is constant except Z. Therefore, the delta ray density ($N_\delta$) is proportional to the square of the particle's charge ($Z^2$).

Development of the method for determining the particle charge in the detector requires a calibration curve. It has to be based on the measured characteristics of tracks produced by particles with known charges. Figure 3 depicts the delta-ray density for the known particle's charges with the best fit line yields a slope of 0.133±0.010. We can also evaluate empirically, the constant for particular counting convention for the particle of known charges.

By using above mention methods, we know the charge of some of the cross checked fragments. So, we can easily calculate the number of the delta ray density of those charges and we can use those charges for calibration part, and the charge of the other relativistic charged particles can be estimated with good accuracy. During scanning we located several electromagnetic dissociated events and we used the projectile fragments of those events to get the delta ray density of known charge.

According to Tidmen et al [13], grain configurations to be counted as delta-rays, must attain a minimum displacement of 1.5 μm from the core of the track projected on the plane of the emulsion. Dependence of delta ray density on the particle charge up to 10 charge units is shown in Figure 3. For nuclei having charge (Z) > 19, the number of delta rays becomes very large and it is difficult to count their number reliably.

Thus, by using this methods, we can measure the charge of the projectile fragments in the range 9 < Z < 20 and cross check the identity of the lower charge projectile fragments estimated by the other method.

## II (b) Angle measurement of projectile fragments

The angle measurement of projectile fragment (PF) were performed in the narrow forward cone ($\theta_{Lab} \leq 10^0$) [7, 14]. Before starting angle measurement of PFs, the micrometer scale inside the eyepiece is aligned along x – coordinate. When micrometer scale becomes aligned with the help of the grains, we align the incident beam tracks along x– coordinate by rotating the mechanical stage gently. When beam track and micrometer scale become aligned, the interaction vertex is focused at the center of the field of view (at the center of the cross wire fitted in one of the eye piece) of micrometer scale so that the coordinates of interaction vertex are recorded as x = 0, y = 0 and z = $z_Y$ (initial z-value i.e. at focused interaction vertex). Now we shifted the interaction vertex to one end of the x-scale through certain known distance as shown in Figure 4. The two coordinates ($x_T$ and $y_T$) of the segment of the PF are read from the counter display of digitizing encoder and the third coordinate ($z_T$) has been accurately recorded.

The spatial configuration of each event was reconstructed by measuring three set of x, y, and z coordinates separated by at least 50 μm along ±x - direction for the incident beam track and for each PFs. In other words, coordinate method was used and three point measurement on the beam tracks as well as PFs that help us in fitting straight line on the PFs and the beam track. Then we obtain the projected angle of the secondary track ($\theta_p$) in x – y plane (i.e. plane of emulsion) [7]:

$$\theta_p = \tan^{-1}(\Delta y / \Delta x) \qquad (1)$$

The dip angle ($\theta_d$) is given by

$$\theta_d = \tan^{-1}[(\Delta z \times S) / (\Delta x^2 + \Delta y^2)^{1/2}], \qquad (2)$$

Where $\Delta z$ is the change in z- coordinate while travel distance $\Delta x$ and $\Delta y$ in the ( x – y ) plane. S is the shrinkage factor. The space angle ($\theta_s$) is given by

$$\theta_s = \cos^{-1}[\cos \theta_p / \{1 + \tan^2(\theta_d)\}^{1/2}]. \qquad (3)$$

The angle of other type of secondary particle tracks has been measured by a Goniometer attached to the eyepiece tube of the microscope. The goniometer is provided with a vernier scale that can yield angles with an accuracy of $0.25^o$.

## III. Results and discussion

## III(a). $^{84}Kr_{36}$ Projectile Fragments Charge Spectrum up to 10 charge unit

In this analysis, we have adopted some of the above mentioned methods for charge estimation of projectile fragments up to 10 charge units. Kr projectile fragments spectrum was compared with other heavier and lighter projectiles fragments charge spectrum having similar beam energy in figure 5 and in figure 6 charge spectrum of similar projectile (Kr) having variable beam energy was compared. In both figures, spectra are presented up to 10 charge units and after that we clubbed all other heavier fragments and called it as more than 10 charge units. The data points are from the present analysis result of $^{84}Kr_{36}$ at around 1 A GeV and histograms are the results from other experiments. The cross checks of charge estimated with other methods reveals that the obtained charge spectrum has reached accuracy up to ±1 charge unit. Error bar shown in Figure 5 and 6 are the statistical errors.

It is evident from figure 5(a) that the production of single, double and more than 10 charge unit projectile fragments have dependence on projectile mass number. Whereas other charge projectile fragments production shows mixed nature. The minimum and maximum production ranges of projectile fragments for the projectile mass number ranging in between 40 to 238 are as follows: 40-65% (z=1); 7 - ~18% (z=2); ~3-7% (z=3); >1-4% (z=4); 0.8-3% (z=5); 0.4 - ~0.9% (z=6); 0.3-0.9% (z=7); >0.2-0.8% (z=8); >0.1-0.5% (z=9); >0.1-0.3% (z=10) and 4 - ~10.5% (z≥11). Due to large statistical error and very narrow energy range, it is very hard to conclude any dependence of projectile fragments production on kinetic energy of projectile from Figure 5(b).

## III(b). Emission angle distribution of projectile fragment

The quantum mechanical features such as Fermi motion is considered to have influence on the angular distribution of emitted particles. Therefore, it is interesting and also important to study and understand the angular distribution of the projectile fragments emitted in the Interaction of $^{84}Kr_{36}$ with emulsion target at relativistic energy. The normalized projected angle distribution of the identified single, double and multiple (z≥3) charge projectile fragments emitted in the interaction are shown in Figure 6. The distributions are best fitted by the Gaussian function, $f(x) = p^o \times exp(-0.5 \times ((x-p^1)/p^2)^2)$.

The mean emission angle decreases with increase in the charge of the projectile fragments is evident from figure 6. For single charge projectile fragments dispersion of the distribution from the mean, 3.306±0.125, is largest and have value 3.260±0.098 while sigma (mean) value of the fitted function are 2.564±0.114 (2.268±0.147) and 2.434±0.119 (1.532±0.182) for double and multiple

charged projectile fragments, respectively. It also reflects that most (80%) of the multiple charged fragments are emitted within 4º while for similar number of PFs emitting up to 5º and 7.5º for double and single charge PFs, respectively. It can be also seen from Figure 6 that around 10.5%, 7% and 6% PFs having multiple, double and single charge, respectively are emitting at zero degree.

The normalized dip angle distribution of same PF tracks is shown in Figure 7, fitted with above mentioned Gaussian function. From this figure, it is clear that the mean emission dip angle also decreases with the increase in charge of the PFs. For single charge projectile fragments dispersion of the distribution from the mean, 3.301±0.138, is largest and have value 3.409±0.107 while sigma (mean) value of the fitted function are 2.723±0.119 (2.344±0.155) and 2.485±0.135 (1.369±0.209) for double and multiple charged projectile fragments, respectively. It also reflects that most of the multiple charged fragments are emitted within 4º while for similar number of PFs emitting up to 5º and 7º for double and single charge PFs, respectively. It can be also seen from Figure 7 that around 11%, 7.5% and 6% PFs having multiple, double and single charge, respectively are emitting at zero degree.

The normalized space angle distribution of same projectile fragment tracks is shown in Figure 8 fitted with above mentioned Gaussian function. From this figure, it can be seen that the mean emission space angle also shows similar trends as shown by projected and dip angle distributions i.e. mean emission space angle decreases with increase in the charge of the projectile fragments. For single charge projectile fragments dispersion of the distribution from the mean, 3.062±0.115, is largest and have value 2.676±0.216 while sigma (mean) value of the fitted function are 2.515±0.694 (2.303±0.302) and 1.380±0.409 (1.371±0.159) for double and multiple charged projectile fragments, respectively. It also reflects that most of the multiple charged fragments are emitted within 4º while for similar number of PFs emitting up to ~6º and ~7º for double and single charge PFs, respectively. It can be also seen from Figure 8 that around 10.5%, 7.5% and 5.5% PFs having multiple, double and single charge, respectively are emitting at zero degree.

Therefore from figures 6, 7 & 8 we can infer that, there are no significant change in the mean and sigma values of particular type of projectile fragments emission angles. It means over all emission shape of three major projectile fragments group is conical.

## III(c). Emission angle distribution based on target species

The variation of emission angle (space angle) of fragments in $^{84}$Kr interactions with individual target group [H, CNO, Ag(Br) and composite emulsion] for single, double and multiple charge PF's were studied and are depicted in the figures 9, 10 and 11 respectively. The target identification and separation method for this experiment is explained in ref. [4]. The experimental data points are represented by symbols and different types of lines (solid, dashed and dotted) are the best fitting function.

Figure 9, 10 and 11 represents normalized distribution of space angle for single, double and multiple charge projectile fragments emitted or decayed during interaction of $^{84}Kr_{36}$ beam with different emulsion detector target groups. 7%, 9% and 9.5% of PF's are emitted at zero angles for H-target, whereas for heaviest target group Ag(Br) 4%, 5.5% and 10.5% were observed for single, double and multiple charge projectile fragments respectively. The maximum values of single charge PF's is 12%, 10% and 8%; for double charge PF's it is 11.5 %, 10% and 9%; and for multiple charge PF's it is 10%, 10.5% and 11% respectively for H, CNO, and Ag(Br) targets. Tailing portion of distribution also follows the similar pattern but in reverse order.

It can also be seen from figure 9, 10 and 11 for single, double and multiple PF's respectively that with the increase in mass number/mean mass number of target group the mean values of fitted function are shifting towards higher emission angle and the dispersion of the distribution is also becoming wider. For a given type of PF's the shapes of all the distributions is similar. These distributions are crossing each other in between 4-5 degree for single charge PF's, at ~4 degree of emission angle for double charge PF's and no one crosses each other in case of multiple charge PF's. This implies that, for single and double charge PF's at the above mentioned angles momentum transfer during interaction is almost similar for all the target species and after that number of PFs having larger momentum is large in case of heavier targets. Whereas for multiple charge PF's, momentum transfer during interactions is showing strong dependence on the mass number of target group throughout the entire emission angles.

The distribution of fitted mean emission angle values are plotted with respect to the charge of the projectile fragments for different target groups including composite emulsion target is shown in figure 12. The figure infers a strong negative dependence of mean emission angle with respect to charge and positive dependence with mass number of target group. Therefore, the mean transverse momentum also follows the similar dependent with the charge of PF and mass number of target. This shows as the degree of breakup of target increases i.e. the impact parameter decreases, a greater fraction of the heavier projectile fragments, alphas and singly charged fragments scatters at larger angles within the forward cone. Since the transfer of momentum in general is larger in case of heavier mass number target.

## III(d). Effects on emission angle due to the charge of associated projectile fragments

From above details it is interesting to study the effects on emission angle for different projectile fragments due to their associated projectile fragments, which are emitted during interaction or decay products of heavy projectile fragments. We performed such study up to the 10 charge unit of projectile fragments and clubbed the charges higher than 10 units together.

We measured the space angle difference ($\Delta\theta_S$) between considered projectile fragments with respect to the rest of the observed projectile fragments in a interaction and plotted the normalized distribution of these difference with respect to charge of the projectile fragments in figure 13. Space angle difference measured considering the PF of particular charge with respect to the rest projectile fragments of event referred to the charge of the considered projectile fragment. The positive and negative signs for angles are just representing upward and downward location of the considered PFs with respect to the beam direction.

Lighter charge projectile fragments are showing larger dispersion from rest of the projectile fragments is depicted in figure 13. It can be seen from Figure 13 that lighter charge ($Z<9$) projectile fragments are showing two peaks, one in positive (upward) and other one is in negative (downward) side. The heavier charge more than 10 charge units just merge and do not show two peaks behavior exhibited by lighter charge projectiles. The ratio of mean value of up and downward peaks distribution is shown in Figure 14. If both side peaks are located at same position i.e. symmetry then the ratio must be at 1. But the best fit of the distribution comes out to be 0.93 which is close to unity. From the figure 14 we can see that the variation of our data points is very close to the expected results. Therefore, proves the symmetrical nature of projectile fragments emission shown in figure 13. It may be concluded from the Figure 14 that the lighter charges are showing symmetry distribution behavior.

It can be seen from Figure 15 that the ratio of the up and downward peak area seems to be equal under certain fix value of sigma. Dotted line is the expected value of ratio and the solid line is showing best fit of distribution is at 1.01±0.04. It may be concluded from Figure 15 that almost equal number of projectile fragments for each charge are symmetrically emitted in interactions.

If we examine carefully, the normalized distribution of the space angle difference ($\Delta\theta_s$) of different charge projectile fragments with respect to the rest of the projectile fragments of the interactions as shown in Figure 13. We can see one small but clear peak on both sides of the big peaks. The ratio of mean and area of the peak are plotted in Figure 16 and 17. For cross check of the symmetrical distribution, we fit data point with best fitting function and found the mean value of each peak. From Figure 16 it can be seen that the best fit line is slightly above (1.02) the expected line at one. It shows that the both small peaks have similar mean value i.e. they are located at the same position but in opposite sides of the beam direction. It also shows that there are certain numbers of projectile fragment of same charge having different emission value difference. It means some projectile fragments have different emission time and therefore it is possible that they are coming from the decay products of the heavier projectile fragments of the interactions.

The ratio of the area under small peaks is plotted in Figure 17. The best fit solid line (0.93) is close to the expected dotted line, showing similar area of the small peaks of all lighter charge within 7 % of the dispersion margin from the expected value. Here we may assume that the similar small peaks are present at other side of the big peaks considering equal distribution. It means there are total four small peaks for each big peak for every lighter charge projectile fragments. On the basis of the Figure 17 we can consider almost equal area of those four peaks with 7 % dispersion margin. From Figure 13, we can calculate that 14.30, 6.67, 8.75, 6.52, 9.12, 10.44, 15.80, 11.05 and 11.14% of charge (Z) equal to 1, 2, 3, 4, 5, 6, 7, 8 and 9, respectively of projectile fragments are not coming from direct interaction i.e. are possibly coming from the decay process of the heavy projectile fragments that are by products of the direct interaction or may be some other process.

## IV. Conclusion

It is quite interesting to study the projectile fragmentation process of heavy ions such as $^{84}$kr projectile. The main conclusions of our experiment are the following:

In the above sections a detailed explanation were given on charge estimation of projectile fragment and angle measurements. From the above study we conclude that the production of heavy and intermediate mass fragments is a function of the size of the fragmenting system as well as the beam energy. Lighter charge and intermediate plus heavy charge projectile fragments such as Z=1 & 2 and > 10 shows strong dependence on mass number of projectiles of similar energies. As charge of the projectile fragments increases their emission chances at zero degree also increases i.e. there are less chance of emission at zero degree of single charge projectile fragment. But lighter charge projectile fragments gaining more transverse momentum than heavier charge one. Our study shows emission distribution of projectile fragments and their transverse momentum has strong dependence on the target mass number.

The angle distribution study of the projectile fragments reveals the nature of fragments and the behavior of fragments on each other during emission that affects the Fermi's motion of the particle. From this study we observed the emitted projectile fragments are strongly affected by the rest of the associated projectile fragments. The distribution of the projectile fragments is showing symmetrical nature for lighter charge projectile fragments and as we move from lower to higher charge symmetrical distribution behavior decrease and both peak merge into a single peak. Therefore, heavy charge projectile fragments moving with nearly same velocity as the incident projectile, with very small deviation in comparison to lighter charge projectile fragment and they are not affected too much by their neighbor projectile fragments. We also observed a small but significant amplitude peaks on both side of the big peak for almost all light charge projectile fragments having different $\Delta\theta$ values. It reflects, there are few percent of projectile fragments that are coming from the decay of heavy projectile fragments or any other process.

## Acknowledgment

Authors are grateful to the staff of the GSI, Germany for exposing the emulsion detector plates.

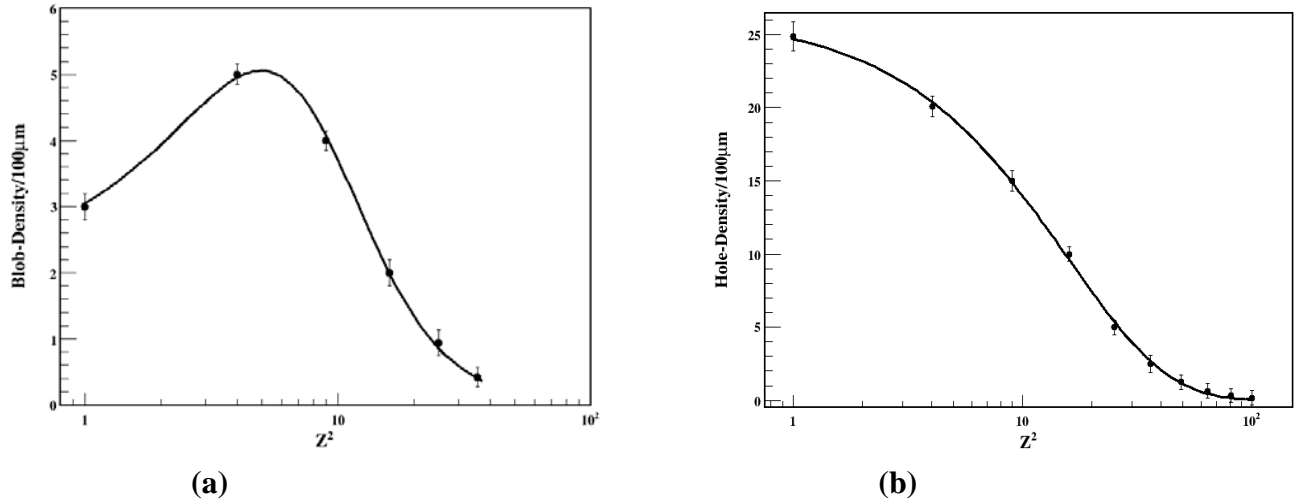

**(a)**                                               **(b)**

**Fig. 1:** Calibration curve of **(a)** Blob's density and **(b)** Hole's density as a function of square of the projectile fragment's charge ($Z^2$). Error bar shown on the data points are pure statistical.

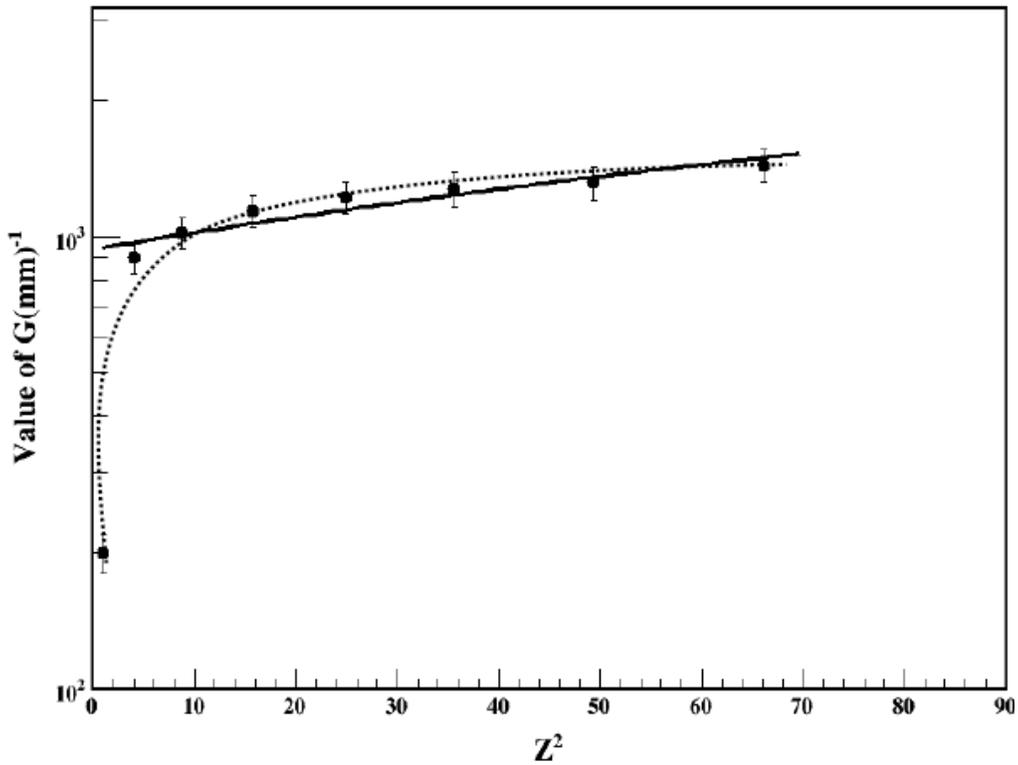

**Fig. 2:** Calibration curve in terms of Gap-length Coefficient as a function of square of the projectile fragment's charge ($Z^2$). The fitting function is first polynomial with slope 8.456±1.792 and intersection at 936.13±53.05.

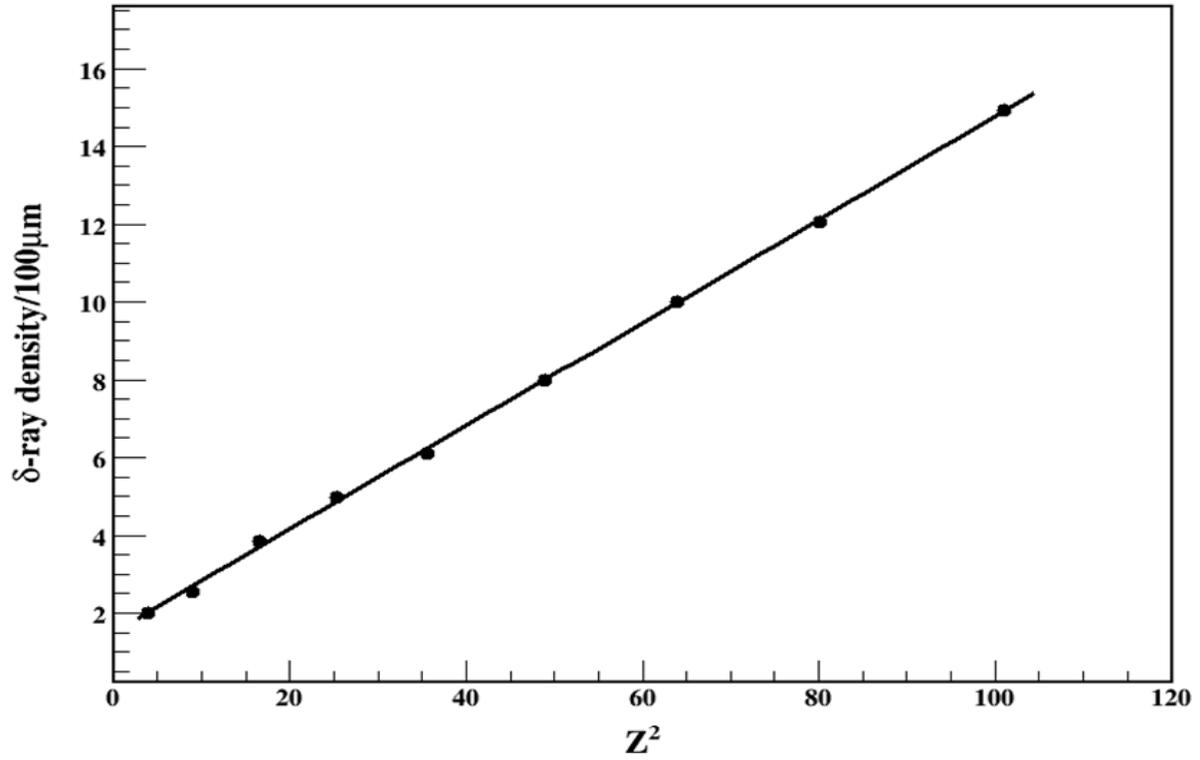

**Fig. 3:** Calibration curve for the little heavier (9 < Z < 20) projectile fragment's charge estimation in terms of delta ray density as a function of $Z^2$.

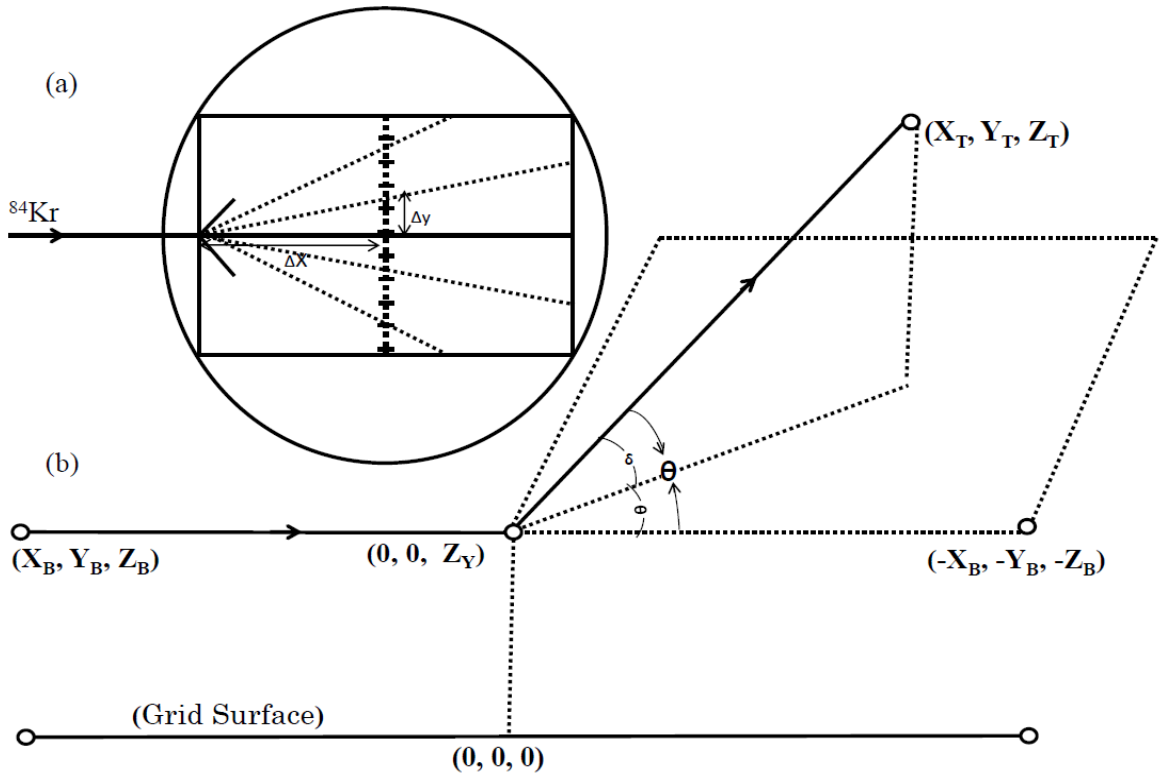

**Fig. 4: (a)** Procedure of the angle measurement and **(b)** definition of coordinates, plans and angles.

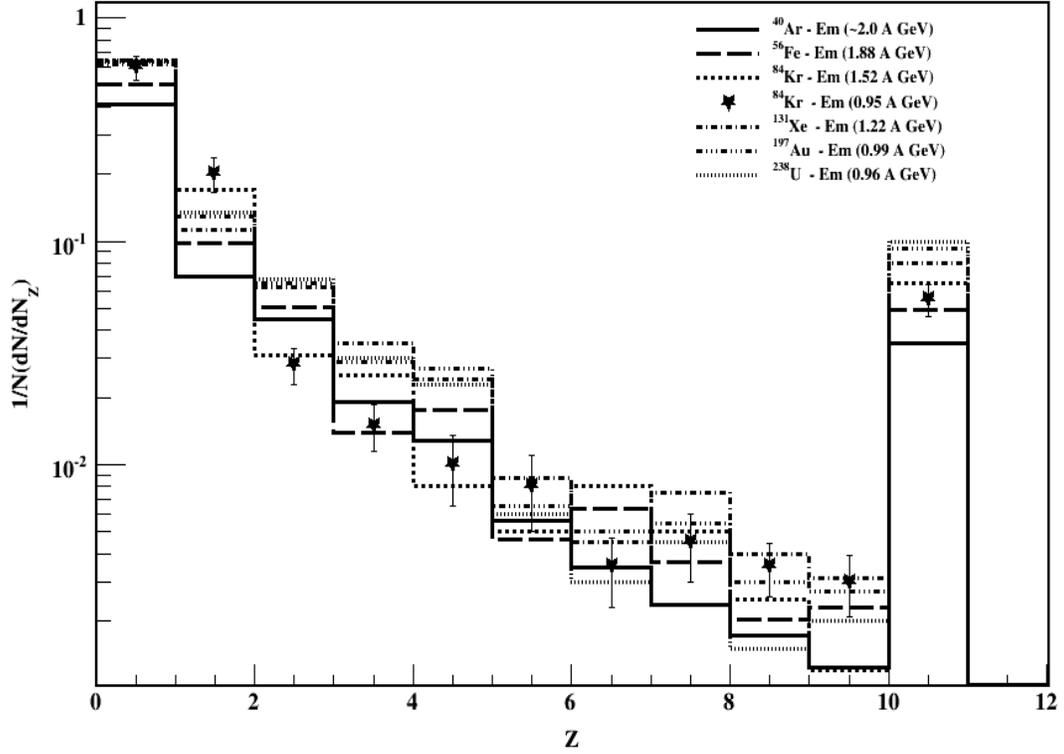

**Fig. 5(a):** Normalized estimated charge spectrum of different projectile at similar energy. $^{84}$Kr at 0.95 [Present work] is compared with the results of $^{40}$Ar at ~2 A GeV [15], $^{56}$Fe at 1.88 A GeV [16], $^{84}$Kr at 1.52 A GeV [17], $^{131}$Xe at 1.22 A GeV [18], $^{197}$Au at 0.99 [19], $^{238}$U at 0.96 A GeV [20].

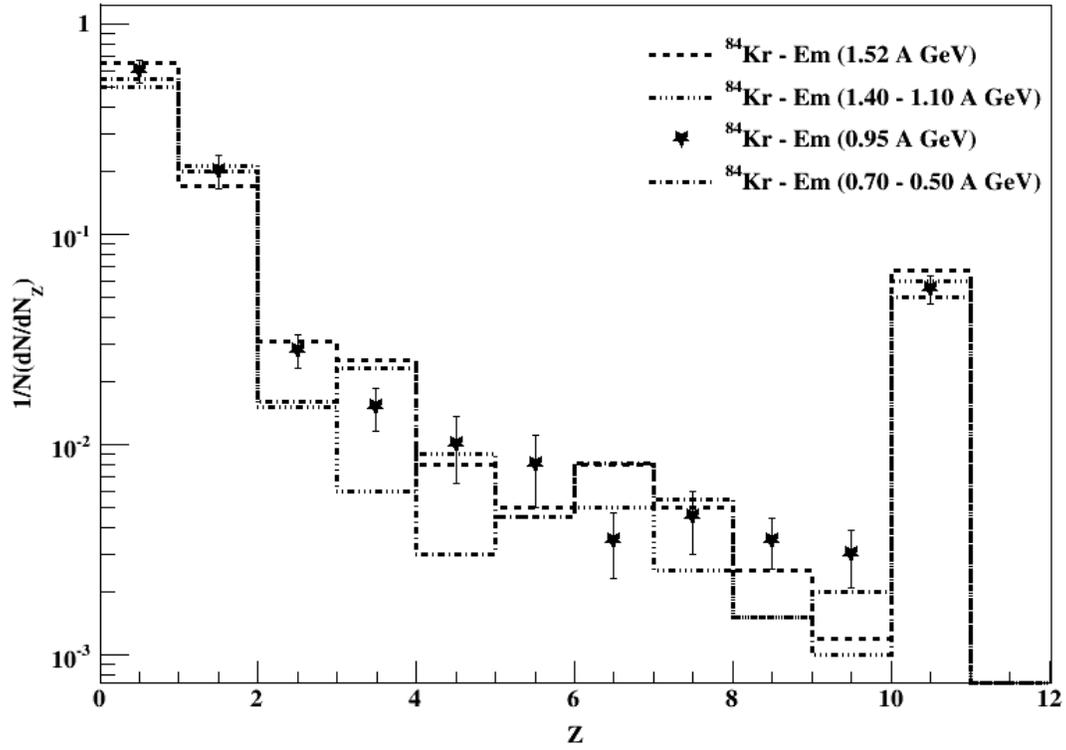

**Fig. 5(b):** Normalized estimated charge spectrum of same projectile with different energy. Data point from $^{84}$Kr at 1.52 A GeV [17], $^{84}$Kr at 1.40 -1.10 [7], $^{84}$Kr at 0.95 A GeV [Present work], $^{84}$Kr at 0.70-0.50 A GeV [7].

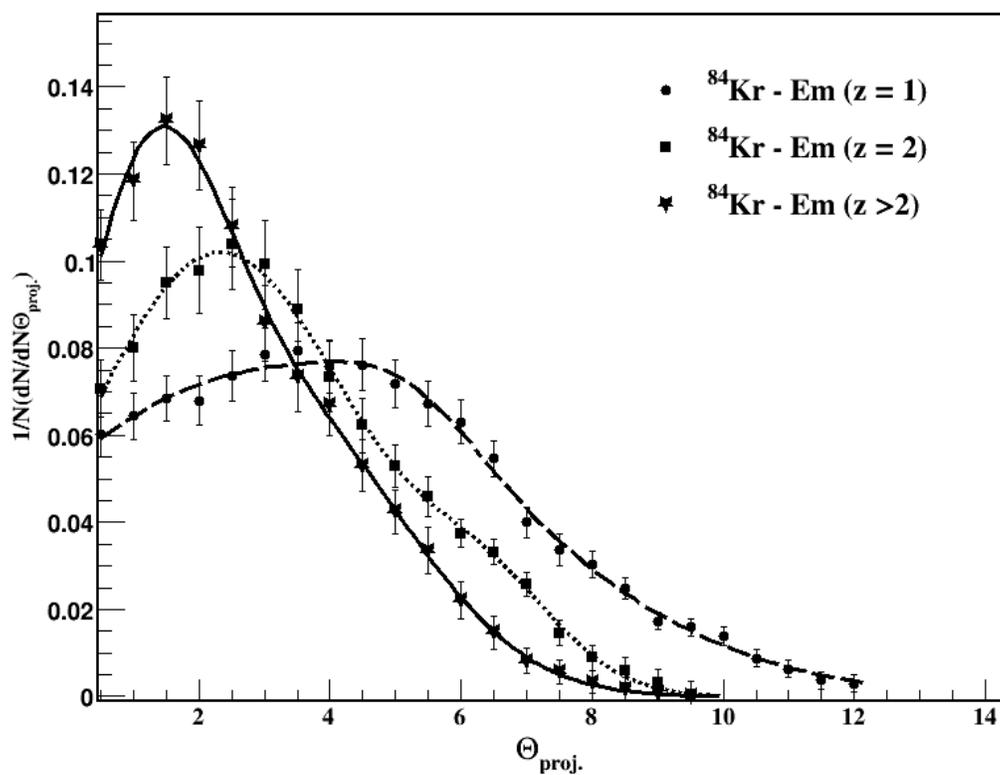

**Fig. 6:** The normalized distribution of projected angle for single, double and multiple charge PFs. The points represent experimental data and solid, dotted and dashed lines are the best fitting function line.

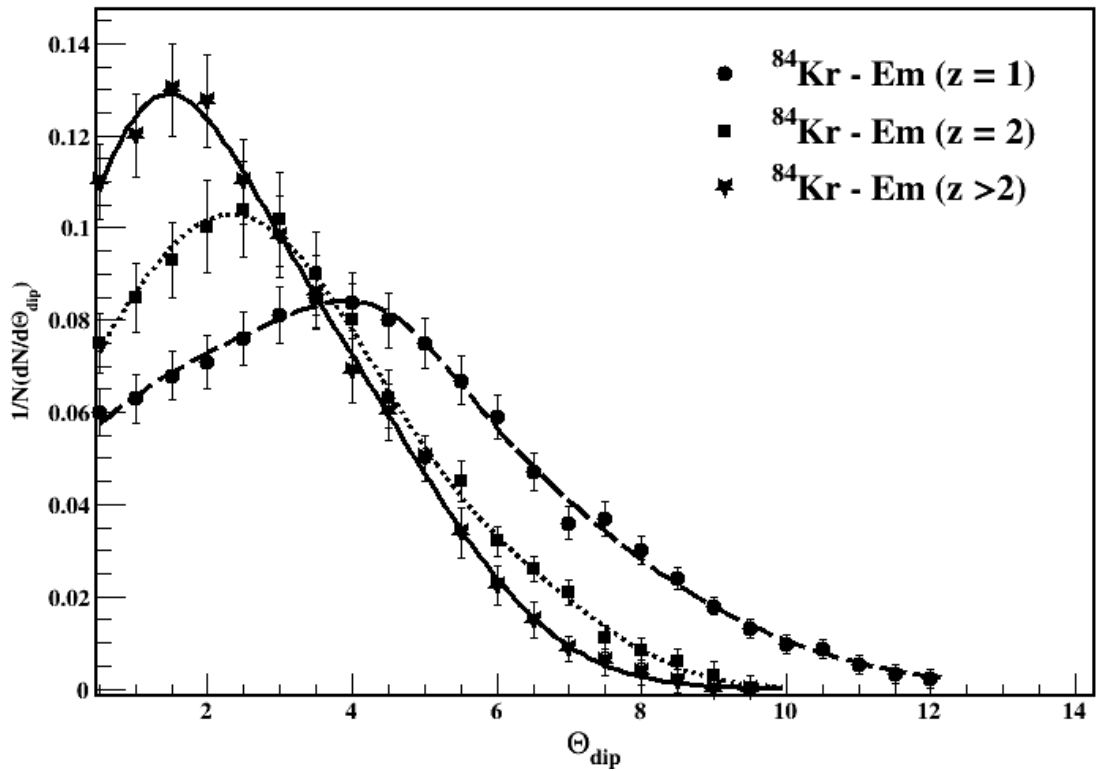

**Fig. 7:** The normalized distribution of dip angle for single, double and multiple charge PFs. The points are representing experimental data and solid, dotted and dashed lines are the best fitting function line.

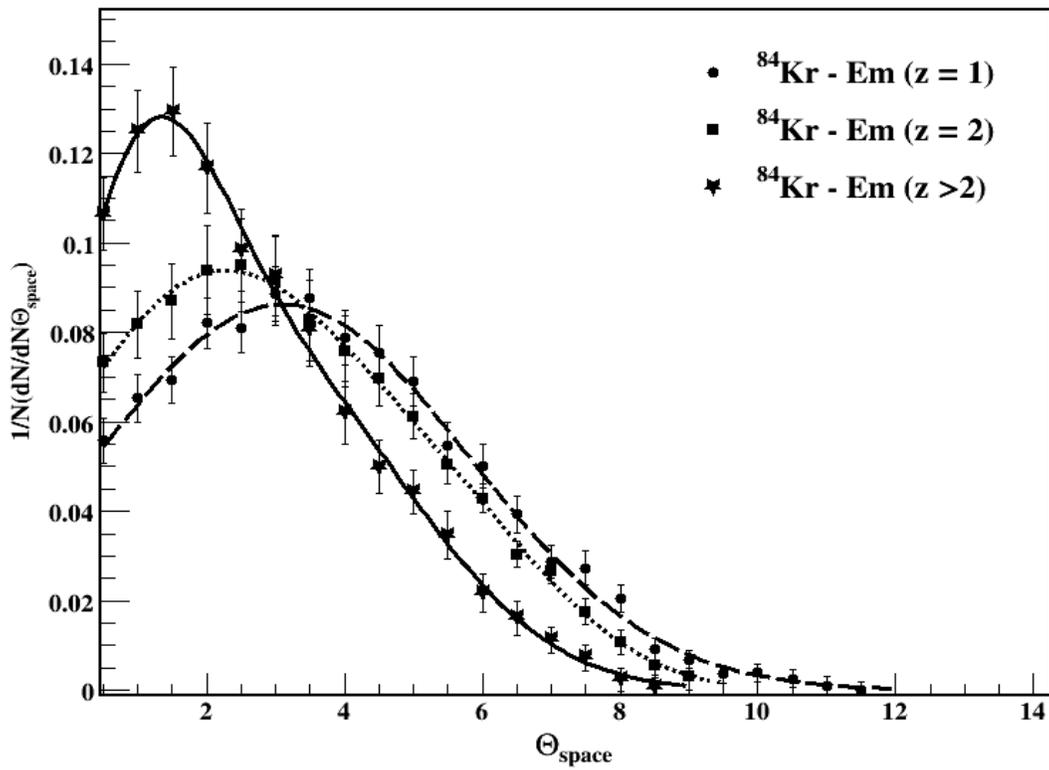

**Fig. 8:** The normalized distribution of space angle for single, double and multiple charge PFs. The points are representing experimental data and solid, dotted and dashed lines are the best fitting function line.

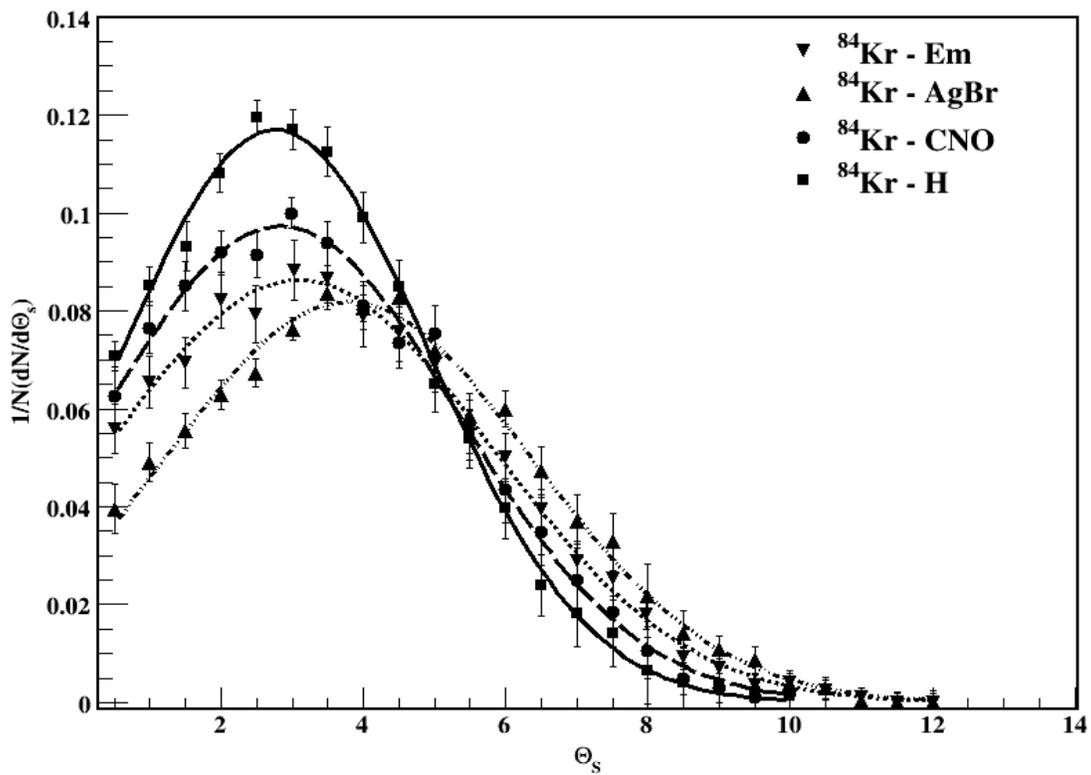

**Fig. 9:** The normalized distribution of space angle for single charge PFs. Points are representing experimental data and solid, dotted and dashed lines are the best fitting function line.

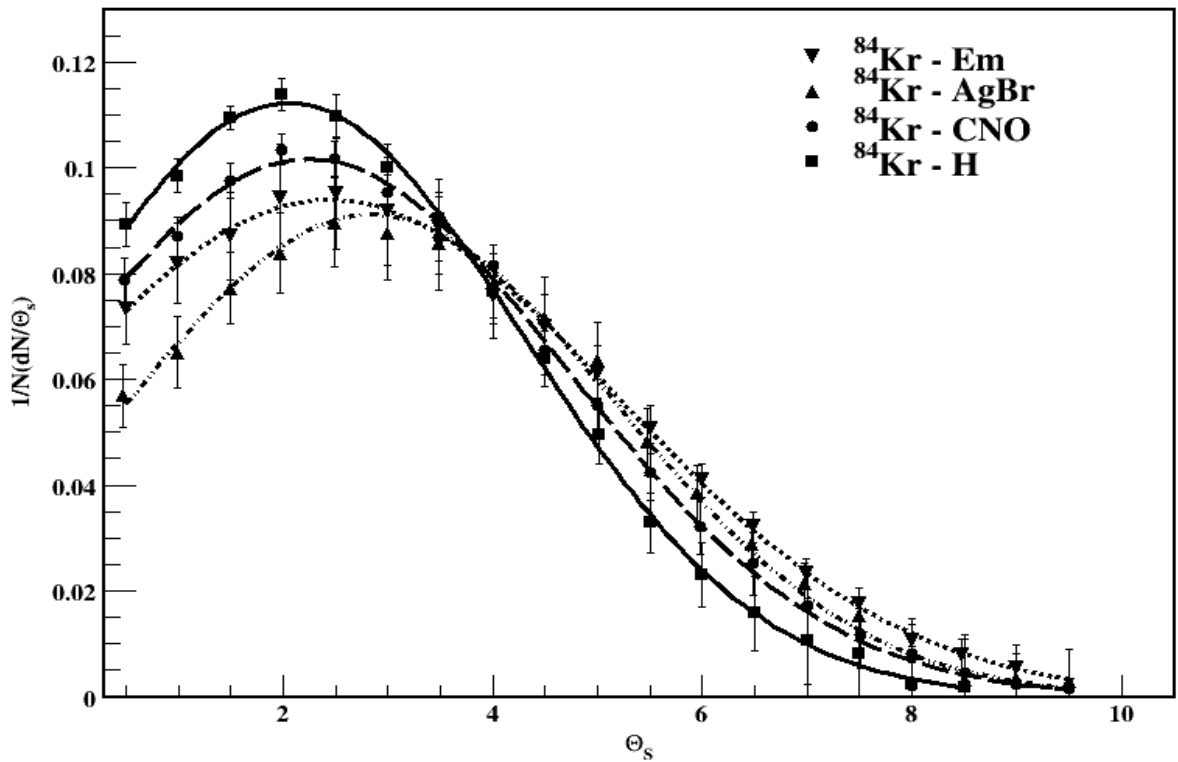

**Fig. 10:** The normalized distribution of space angle for double charge PFs. Points are the experimental data and solid, dotted and dashed lines are the best fitting function line.

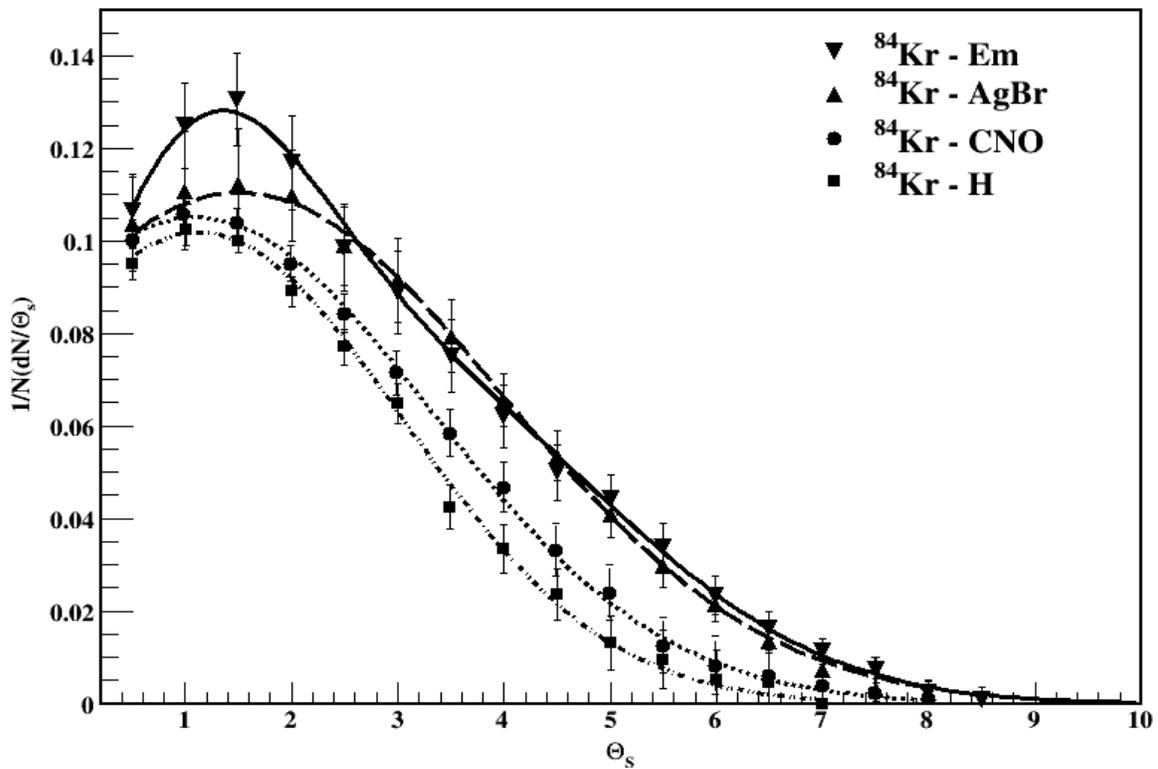

**Fig. 11:** The normalized distribution of space angle for multiple charge projectile fragments. Points are representing the experimental data and solid, dotted and dashed lines are the best fitting function.

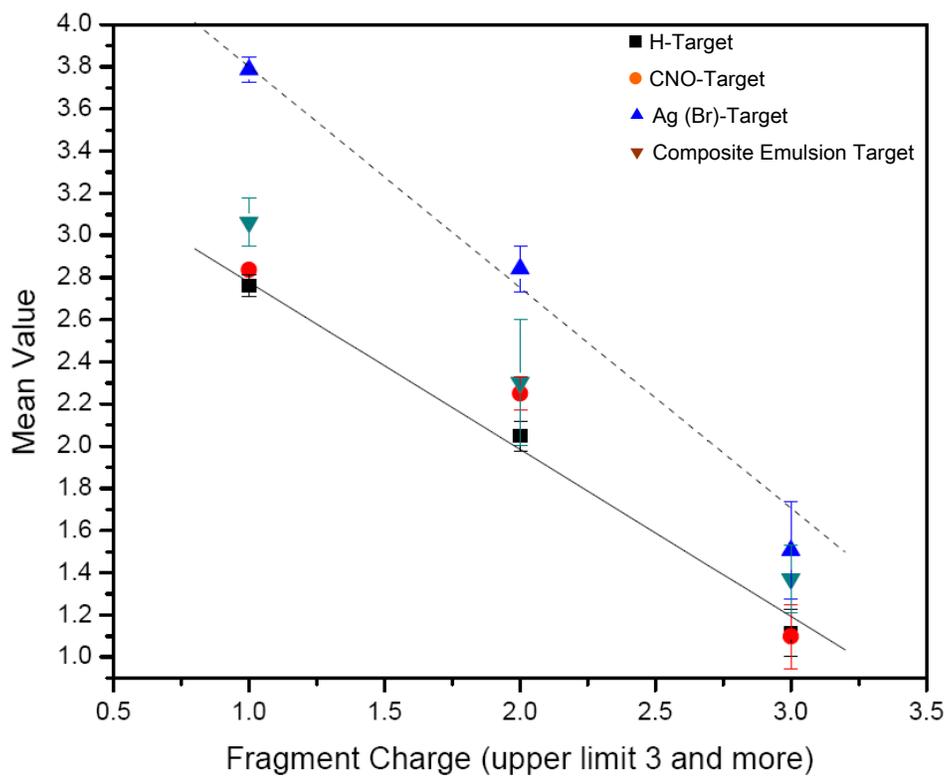
**Fig. 12:** Fitting function's mean value variation with charge of the projectile fragment for different emulsion target groups.

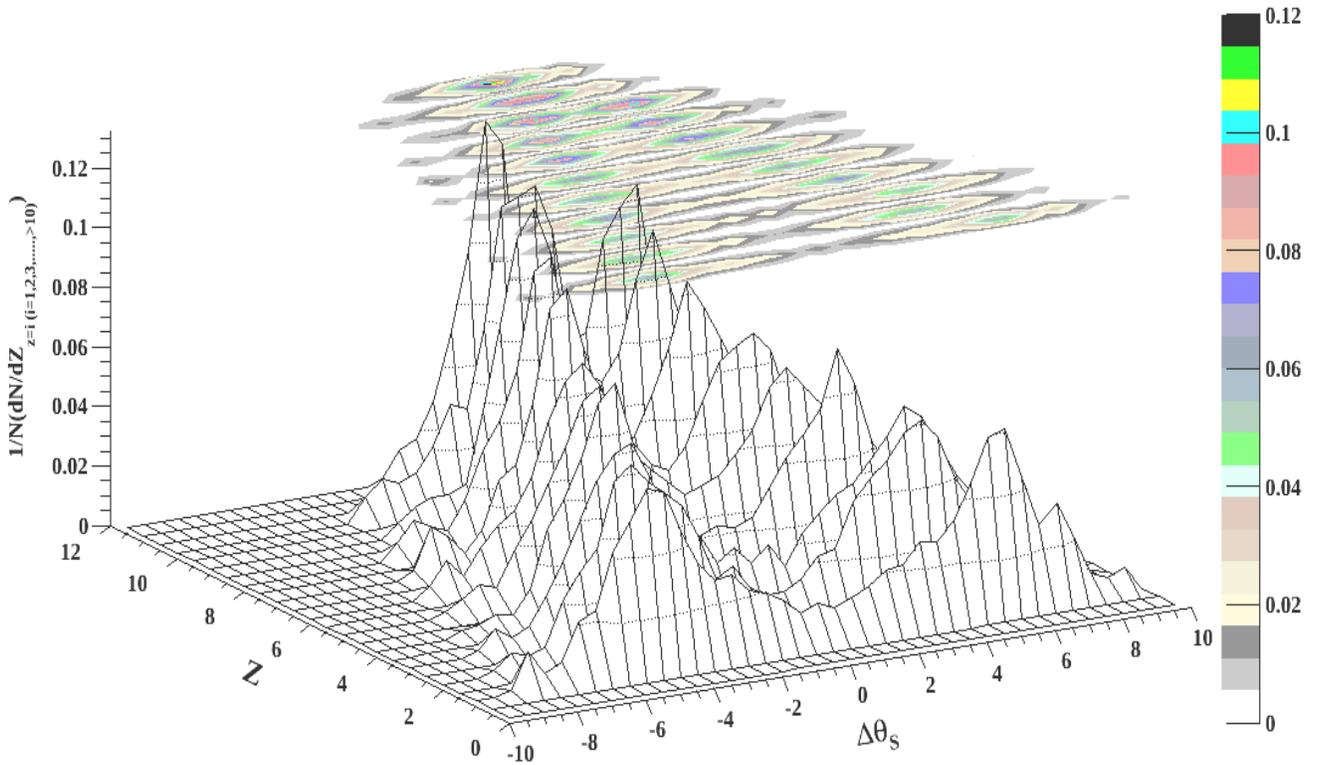

**Fig. 13:** Normalized distribution of the space angle difference (Δθ$_s$) of different charge projectile fragments with respect to the rest of the projectile fragments of the interactions.

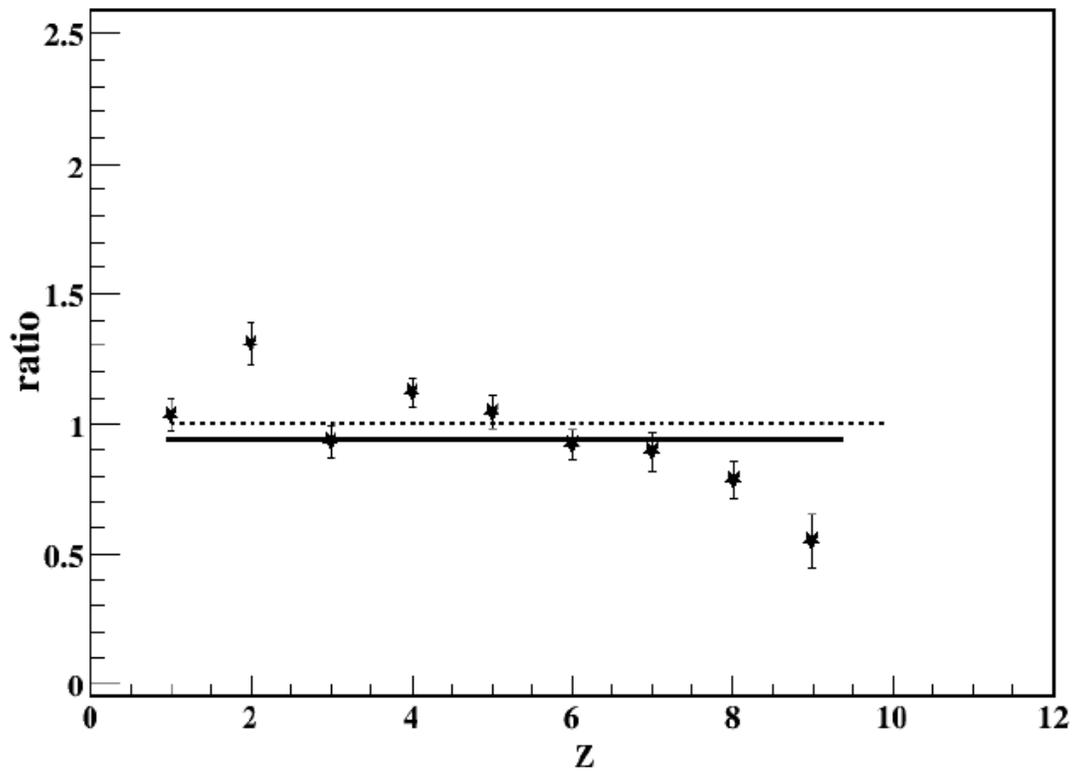

**Fig. 14:** The ratio of mean value of big peaks located up and downward of the beam direction is distributed.

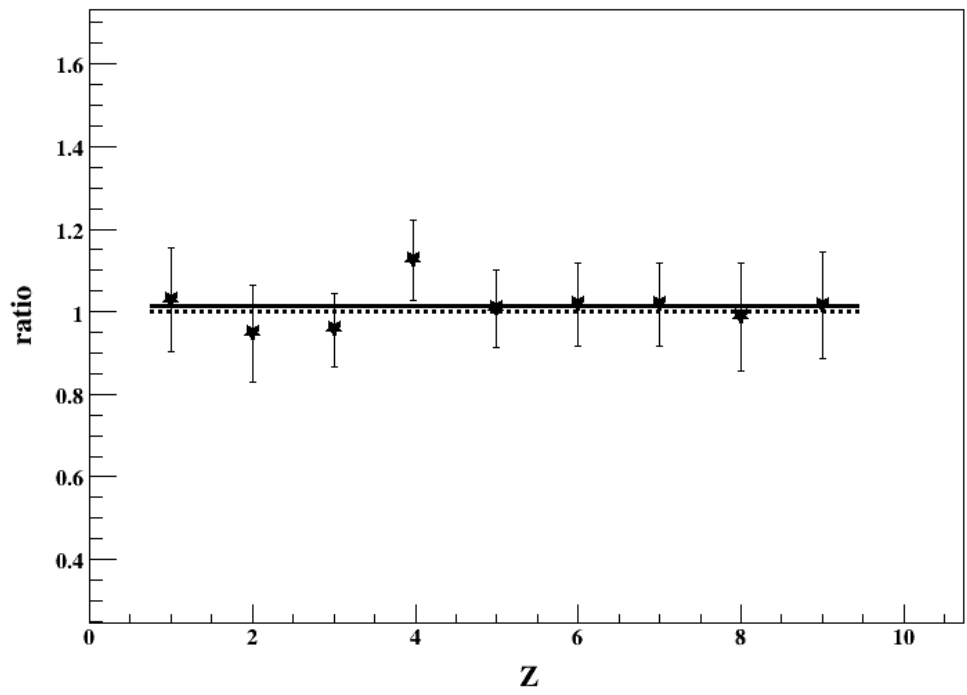

**Fig. 15:** Ratio of the area located under three sigma region of the big peaks located up and downward of the beam direction is distributed.

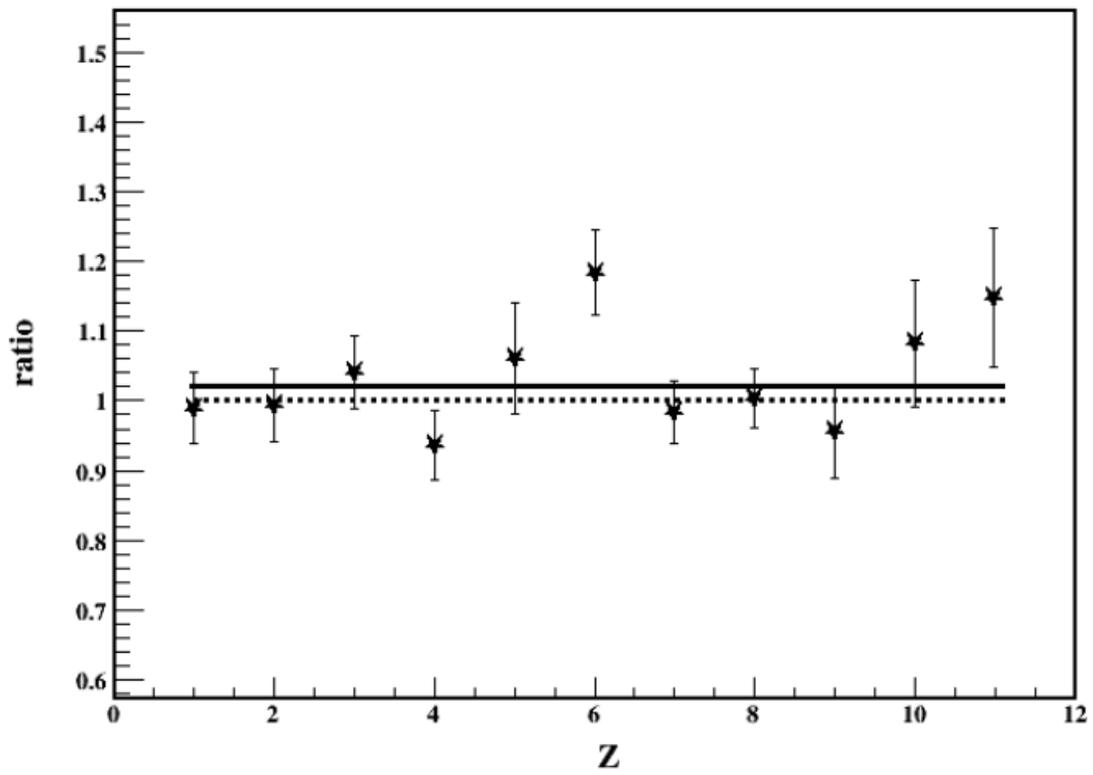

**Fig. 16:** Data point are the ratio of the mean values of small peak and solid line is the best fitting for data point, dotted line are expected value.

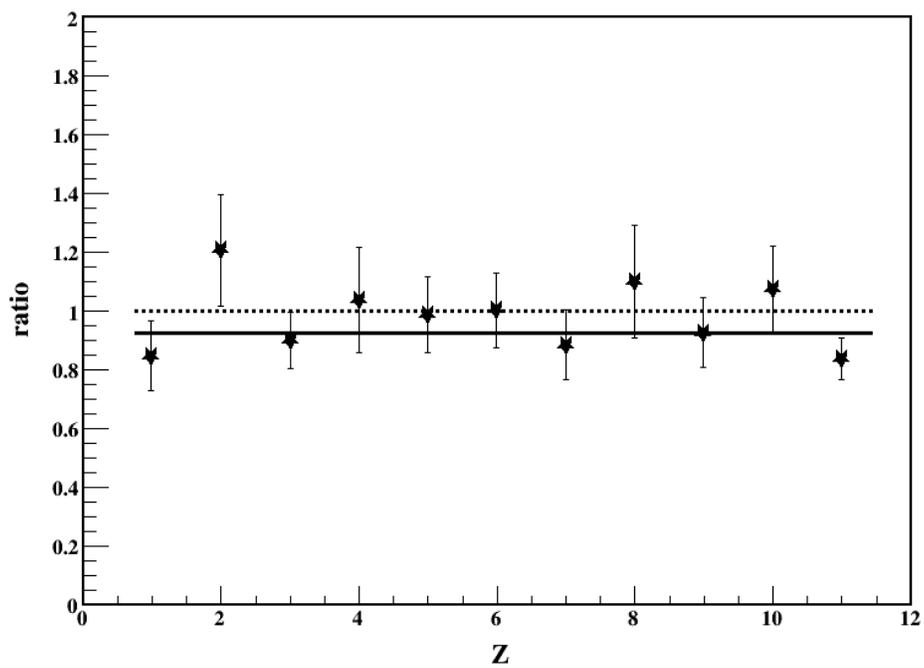

**Fig. 17:** Data points are the ratio of the area under small peak and solid line is the best fitting for data point, dotted line are expected value.